\documentclass[twocolumn,showpacs,preprintnumbers,amsmath,amssymb]{revtex4}
\usepackage{amsmath}
\usepackage{amsmath}
\makeatletter
\newenvironment{tablehere}
{\def\@captype{table}}
\makeatother
\usepackage{amssymb}
\usepackage{txfonts}

\usepackage{graphicx}
\usepackage{bm}
\usepackage{color}

\begin{document}


\setlength{\parindent}{1em}

\title{ Magic Wavelengths for Terahertz Clock Transitions}

\author{Xiaoji Zhou}\thanks{Electronic address: xjzhou@pku.edu.cn }
\author{Xia Xu}
\author{Xuzong Chen}
\author{Jingbiao Chen}\thanks{Electronic address: jbchen@pku.edu.cn }

\affiliation{ School of Electronics Engineering $\&$ Computer
Science, Peking University, Beijing 100871, P. R. China}
%



\begin{abstract}
Magic wavelengths for laser trapping of boson isotopes of
alkaline-earth Sr, Ca and Mg atoms are investigated while
considering terahertz clock transitions between the $^{3}P_{0},
^{3}P_{1}, ^{3}P_{2}$ metastable triplet states. Our calculation
shows that magic wavelengths of trapping laser do exist. This result
is important because those metastable states have already been used
to realize accurate clocks in the terahertz frequency domain.
Detailed discussions for magic wavelength for terahertz clock
transitions are given in this paper.

\end{abstract}


\maketitle

\section{Introduction}

Frequency standards have achieved an unprecedented success in
experimental demonstrating accuracies of $4\times10^{-16}$ with a
cesium microwave fountain clock ~\cite{Santarelli} and
$1.9\times10^{-17}$ with an ion optical clock~\cite{Schneider,
Rosenband}. For optical frequency standards based on neutral atoms,
in order to effectively increase the interrogation time, Katori
proposed to utilize optical lattice trap formed with a magic
wavelength trapping laser~\cite{Katori,Takamoto}. This clever
technique greatly enhanced established high-accuracy optical
frequency standard with neutral Sr atom to an accuracy of $1
\times10^{-16}$ ~\cite{Ye,Blatt,Ludlow}. Different optical clock
schemes based on Ca~\cite{Wilpers}, Yb~\cite{Barbaer,Barber} atoms
trapped with magic wavelength lasers have been proposed.

Optical trap with a far off-resonant laser is a very useful tool for
the confinement of cold atoms. Nevertheless, for the precision of
clock transitions in frequency standards, light shift due to
trapping laser has to be avoided. Thus the wavelength of the
trapping laser should be tuned to a region where the light shift for
the clock transition is eliminated, that means the light shifts of
the two clock transition states cancel each other. The wavelength
$\lambda$ is called magic wavelength~\cite{Katori,Takamoto}.
Recently, cesium primary frequency standard with atoms trapped in an
optical lattice with a magic wavelength was
suggested~\cite{Flambaum,CPL}, and possible magic wavelengths for
clock transitions in aluminium and gallium atoms were also
calculated~\cite{Beloy}.

In contrast to the above mentioned magic wavelengths for optical
clock transitions and microwave clock transitions, here we
investigate magic wavelengths for terahertz clock transitions.
Absolute frequency standards in the terahertz domain with fine
structure transition lines of the Mg and Ca metastable triplet
states were first proposed in 1972 by Strumia~\cite{Strumia}. After
more than twenty years of continuing improvement, a frequency
standard based on the $^{3}P_{1} - ^{3}P_{0}$ Mg transition and
thermal atoms in a beam has reached an uncertainty of
$1\times10^{-12}$ ~\cite{Godone, Levi}. However, these potential
terahertz transitions for high-resolution clock references have
never been experimentally investigated with laser cooled or laser
trapped atoms.

In this paper, we present our most recent calculation of trapping
laser magic wavelengths for Sr, Ca and Mg atoms, considering
different possible clock transitions between metastable triplet
states $^{3}P$. Accurate terahertz  clocks could then be built based
on such atoms which are cooled and trapped in an optical lattice.

\section{Theoretical description}

\begin{figure}[t] \centering
\includegraphics[width=10cm]{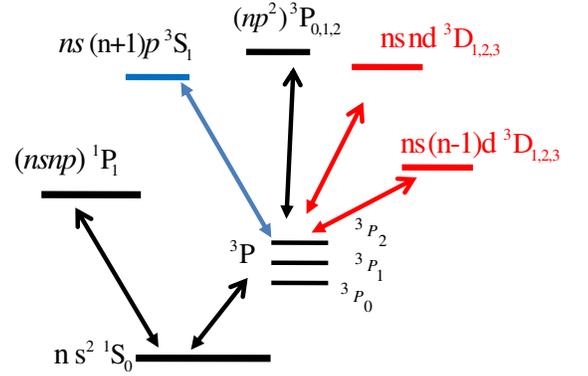}
 \caption{(Color online) Simplified diagram of the lowest energy
levels for alkaline-earth atoms and some possible laser couplings. }
\end{figure}

For alkaline-earth atoms, two valence electrons result in two series
of atomic energy levels as the electron spins can be parallel
(triplet states) or anti parallel (singlet state). The energy
diagram can be simplified as shown in Fig. 1. The ground state is
$^{1}S_{0}$, and the lowest excited states $nsnp$ are $^{1}P_{1}$
and $^{3}P_{J}$ which can be divided into three fine structure
sublevels $^{3}P_{2}$,$^{3}P_{1}$,$^{3}P_{0}$. For the $^{3}P_{J}$
state, transitions to higher states can be divided into three groups
: $^{3}P_{J}-^{3}S_{1}$,$^{3}P_{J}-^{3}P_{J}$ and
$^{3}P_{J}-^{3}D_{J}$.

By second-order perturbation theory, the energy shift
$U_{i}(\omega,p,m_i)$ of atomic state $|i\rangle$ with energy
$E_{i}$ and Zeeman sublevel $m_{i}$, which is induced by a trapping
laser field with frequency $\nu$=$\omega/2\pi$, polarization $p$,
and irradiance intensity  $I$ can be expressed as
$U_i(\omega,p,m_i)$=$-\alpha_i(\omega,p,m_i)I/2\epsilon_0 c$ with
the induced polarizability $\alpha_i$.

The polarizability can be calculated by summing up the contributions
from all dipole interactions between the fine structure state
$|i\rangle$ and $|k\rangle$ with the Einstein coefficient $A_{Jki}$
(spontaneous emission rate for $E_k$$>$$E_i$), Zeeman sublevels
$m_i, m_k$ and transition frequency
$\nu_{Jki}$=$\omega_{Jki}/2\pi$~\cite{grimm, Carsten},
\begin{eqnarray}
\alpha_i=6\pi c^3\epsilon_0
\sum\limits_{k,m_k}\frac{A_{Jki}(2J_k+1)}{\omega^2_{Jki}(\omega^2_{Jki}-\omega^2)}\left(\begin{array}{ccc}
J_i & 1 & J_k\\m_i &p &-m_k\end{array}\right)^2
\end{eqnarray}
where
\begin{eqnarray}
A_{Jki} = \frac{e^2}{4\pi\epsilon_0}\frac{{4\omega
_{Jki}^3}}{{3\hbar {c^3}}}\frac{1}{{2{J_k} + 1}}|\left\langle
{{\beta_k}{J_k}} \left\| D \right\| {{\beta _i}{J_i}}
\right\rangle|^2
\end{eqnarray}
Here e is the electron charge, $\hbar\omega_{Jki}$ is the energy
difference between fine structure states $|k\rangle$ and
$|i\rangle$, $\beta$ denotes other quantum numbers of the state, and
${\left\langle {{\beta_k}{J_k}} \left\| D \right\| {{\beta_i}{J_i}}
\right\rangle } $ is the dipole reduced matrix element. The
expression in large parentheses in Eq.(1) denotes a $3J$ symbol
which describes the selection rules and relative strength of the
transition depending on the involved angular momenta $J$, the
projection $m$, and the polarization $p$.

If we know $\omega_{Jki}$ and $A_{Jki}$ in Eq.(1), we can get the
polarizability $\alpha_i$. However, typically the literature gives
the total transition rate $A_T$ from a given excited state to the
fine structure manifold states below. So we need establish the
relation 

\begin{tablehere}
\begin{scriptsize}
\begin{widetext}
\caption{Sr element: Transition Wavenumbers (WN)($cm^{-1}$)
corresponding to $\omega_{Jki}$, Einstein Coefficients for fine
structure states $A_J(\times10^6s^{-1})$ and Total
$A_T(\times10^6s^{-1})$ for fine structure states manifold,
Correction Factors $\boldsymbol{\zeta}$. The Wavenumber data
originate from~\cite{EMoore}.}
\begin{tabular}{|c|c|c|c|c|}\hline
\hline{$$}&\multicolumn{3}{c|}{{$5s^{2}$$^{1}S_{0}$}}&{$$}\\\hline
$\textbf{\textrm{UpperState}}$&{WN}&{${A_{J}}$}&{$\boldsymbol{\zeta}$}&\raisebox{0ex}[0pt]{${A_{T}}$}\\\hline
$5s5p^{1}P_{1}$    &21698.48 &190.01 &1.000                &190.01$^{b}$\\
$5s6p^{1}P_{1}$    &34098.44 &1.87   &1.000                &1.87$^{d}$\\
$5s7p^{1}P_{1}$    &38906.90 &5.32   &1.000                &5.32$^{d}$\\
$5s8p^{1}P_{1}$    &41172.15 &14.9   &1.000                &14.9$^{a}$\\
$4d5p^{1}P_{1}$    &41184.47 &12     &1.000                &12$^{a}$\\
$5s9p^{1}P_{1}$    &42462.36 &11.6   &1.000                &11.6$^{a}$\\
$5s10p^{1}P_{1}$   &43327.94 &7.6    &1.000                &7.6$^{a}$\\
$5s11p^{1}P_{1}$   &43938.26 &4.88   &1.000
&4.88$^{a}$\\\hline
\end{tabular}
\\
\begin{tabular}{|c|c|c|c|c|c|c|c|c|c|c|}\hline
\hline{$$}&\multicolumn{3}{c|}{{$5s5p^{3}P_{0}$}}
&\multicolumn{3}{c|}{{$5s5p^{3}P_{1}$}}
&\multicolumn{3}{c|}{{$5s5p^{3}P_{2}$}} &{$$}\\\hline
$\textbf{\textrm{UpperState}}$&{WN}
&{${A_{J}}$}&{$\boldsymbol{\zeta}$}&{WN}&{${A_{J}}$}&{$\boldsymbol{\zeta}$}&{WN}
&{${A_{J}}$}&{$\boldsymbol{\zeta}$}&\raisebox{0ex}[0pt]{${A_{T}}$}\\\hline
$5s6s^{3}S_{1}$    &14721.275&10.226&1.0828 &14534.444 &29.526 &1.0421 &14140.232 &45.314 &0.9596&85$^{c}$\\
$5s7s^{3}S_{1}$    &23107.193&1.402 &1.0517 &22920.362 &4.106  &1.0264 &22526.15  &6.495  &0.9743&12$^{e}$\\
$5s8s^{3}S_{1}$    &26443.920&0.954 &1.0450 &26257.089 &2.803  &1.0230 &25862.877 &4.464  &0.9776&8.22$^{a}$\\
$5s9s^{3}S_{1}$    &28133.680&0.525 &1.0422 &27946.849 &1.543  &1.0216 &27552.637 &2.464  &0.9790&4.53$^{a}$\\
$5s10s^{3}S_{1}$   &29110.080&0.320 &1.0408 &28923.249 &0.943
&1.0208 &28529.037 &1.508  &0.9797&2.77$^{a}$\\\hline
$5p^{2}$$^{3}P_{0}$&-------  &0.000 &----   &20689.119 &117.64 &0.9803 &-------   &0.000  &----  &120$^{e}$\\
$5p^{2}$$^{3}P_{1}$&21082.618&41.492&1.0373 &20895.787 &30.297 &1.0099 &20501.575 &47.690 &0.9538&120$^{e}$\\
$5p^{2}$$^{3}P_{2}$&-------  &0.000 &----   &21170.317 &31.509
&1.0503 &20776.105 &89.343 &0.9927&120$^{e}$\\\hline
$5s4d^{3}D_{1}$    &3841.536 &0.290 &1.2660 &3654.705  &0.187  &1.0901 &3260.493  &0.009  &0.7740&0.412$^{f}$\\
$5s4d^{3}D_{2}$    &-------  &0.000 &----   &3714.444  &0.354  &1.1444 &3320.232  &0.084  &0.8174&0.412$^{f}$\\
$5s4d^{3}D_{3}$    &-------  &0.000 &----   &-------   &0.000  &----   &3420.704  &0.368  &0.8938&0.412$^{f}$\\
$5s5d^{3}D_{1}$    &20689.423&35.732&1.0544 &20502.592 &26.080 &1.0261 &20108.38  &1.640  &0.9681&61$^{g}$\\
$5s5d^{3}D_{2}$    &-------  &0.000 &----   &20517.664 &47.049 &1.0284 &20123.452 &14.796 &0.9702&61$^{g}$\\
$5s5d^{3}D_{3}$    &-------  &0.000 &----   &-------   &0.000  &----   &20146.492 &59.390 &0.9736&61$^{g}$\\
$5s6d^{3}D_{1}$    &25368.383&14.303&1.0457 &25181.552 &10.492 &1.0228 &24787.34  &0.667  &0.9755&24.62$^{f}$\\
$5s6d^{3}D_{2}$    &-------  &0.000 &----   &25186.488 &18.897 &1.0234 &24792.276 &6.008  &0.9761&24.62$^{f}$\\
$5s6d^{3}D_{3}$    &-------  &0.000 &----   &-------   &0.000  &----   &24804.591 &24.069 &0.9776&24.62$^{f}$\\
$5s7d^{3}D_{1}$    &27546.88 &8.223 &1.0424 &27360.049 &6.043  &1.0213 &26965.837 &0.386  &0.9778&14.2$^{a}$\\
$5s7d^{3}D_{2}$    &-------  &0.000 &----   &27364.969 &10.883 &1.0219 &26970.757 &3.473  &0.9783&14.2$^{a}$\\
$5s7d^{3}D_{3}$    &-------  &0.000 &----   &-------   &0.000  &----   &26976.337 &13.902 &0.9790&14.2$^{a}$\\
$5s8d^{3}D_{1}$    &28749.18 &4.920 &1.0407 &28562.349 &3.619  &1.0206 &28168.137 &0.231  &0.9789&8.51$^{a}$\\
$5s8d^{3}D_{2}$    &-------  &0.000 &----   &28565.959 &6.517  &1.0210 &28171.747 &2.083  &0.9793&8.51$^{a}$\\
$5s8d^{3}D_{3}$    &-------  &0.000 &----   &-------   &0.000  &----   &28176.207 &8.337  &0.9797&8.51$^{a}$\\
$5s9d^{3}D_{1}$    &29490.28 &3.184 &1.0400 &29303.449 &2.342  &1.0203 &28909.237 &0.150  &0.9797&5.51$^{a}$\\
$5s9d^{3}D_{2}$    &-------  &0.000 &----   &29303.449 &4.216  &1.0203 &28909.237 &1.350  &0.9797&5.51$^{a}$\\
$5s9d^{3}D_{3}$    &-------  &0.000 &----   &-------   &0.000  &----
&28914.037 &5.401  &0.9802&5.51$^{a}$\\\hline
\multicolumn{11}{|l|}{$^{a}$ ~\cite{SrA}, $^{b}$ ~\cite{Srb}, $^{c}$
~\cite{Src}, $^{d}$ ~\cite{Srd}, $^{e}$ ~\cite{Sre}, $^{f}$
~\cite{Srf}, $^{g}$ ~\cite{Srg}.}\\\hline
\end{tabular}
\end{widetext}
\end{scriptsize}
\end{tablehere}

\begin{tablehere}
\begin{scriptsize}
\begin{widetext}
\caption{Ca element: Transition wavenumbers (WN)($cm^{-1}$)
corresponding to $\omega_{Jki}$, Einstein Coefficients for fine
structure states $A_J(\times10^6s^{-1})$ and Total
$A_T(\times10^6s^{-1})$ for fine structure states manifold,
Correction Factors $\boldsymbol{\zeta}$. Besides the updated values
listed in Ref.~\cite{Carsten} for $A_T$, the others originate
from~\cite{NIST}.}
\begin{tabular}{|c|c|c|c|c|}\hline
\hline{$$}&\multicolumn{3}{c|}{{$4s^{2}$$^{1}S_{0}$}}&{$$}\\\hline
$\textbf{\textrm{UpperState}}$&{WN}&{${A_{J}}$}&{$\boldsymbol{\zeta}$}&\raisebox{0ex}[0pt]{${A_{T}}$}\\\hline
$4s4p^{1}P_{1}$    &23652.304&218  &1.000                &218 \\
$4s5p^{1}P_{1}$    &36731.615&0.27 &1.000                &0.27\\
$4s6p^{1}P_{1}$    &41679.008&16.7 &1.000                &16.7\\
$4snp^{1}P_{1}$    &43933.477&30.1 &1.000                &30.1\\
$4s7p^{1}P_{1}$    &45425.358&15.3 &1.000                &15.3\\
$4s8p^{1}P_{1}$    &46479.813&6.1  &1.000                &6.1\\
\hline
\end{tabular}
\\
\begin{tabular}{|c|c|c|c|c|c|c|c|c|c|c|}\hline
\hline{$$}&\multicolumn{3}{c|}{{$4s4p^{3}P_{0}$}}
&\multicolumn{3}{c|}{{$4s4p^{3}P_{1}$}}
&\multicolumn{3}{c|}{{$4s4p^{3}P_{2}$}} &{$$}\\\hline
$\textbf{\textrm{UpperState}}$&{WN}
&{${A_{J}}$}&{$\boldsymbol{\zeta}$}&{WN}&{${A_{J}}$}&{$\boldsymbol{\zeta}$}&{WN}
&{${A_{J}}$}&{$\boldsymbol{\zeta}$}&\raisebox{0ex}[0pt]{${A_{T}}$}\\\hline
$4s5s^{3}S_{1}$    &16381.594&9.855 &1.0195 &16329.432 &29.278   &1.0096 &16223.552 &47.845   &0.9899   &87$^{a}$\\
$4s6s^{3}S_{1}$    &25316.340&3.488 &1.0126 &25264.178 &10.397   &1.0062 &25158.298 &17.110   &0.9935   &31\\
$4s7s^{3}S_{1}$    &28822.866&1.573 &1.0110 &28770.704 &4.692    &1.0054 &28664.824 &7.733    &0.9943   &14\\
$4s8s^{3}S_{1}$    &30580.783&1.033 &1.0102 &30528.621 &3.083    &1.0052 &30422.741 &5.084    &0.9947   &9.2\\
$4s9s^{3}S_{1}$    &31590.382&0.606 &1.0099 &31538.220 &1.809    &1.0050 &31432.340 &2.985    &0.9950   &5.4\\
$4s10s^{3}S_{1}$   &32224.147&0.370 &1.0097 &32171.985 &1.105
&1.0049 &32066.105 &1.824    &0.9951   &3.3\\\hline
$4p^{2}$$^{3}P_{0}$&------   &0.000 &----   &23207.480 &179.046  &0.9947 &-----     &0.000    &----     &180\\
$4p^{2}$$^{3}P_{1}$&23306.907&60.474&1.0079 &23254.745 &45.045   &1.0010 &23148.865 &74.033   &0.9871   &180\\
$4p^{2}$$^{3}P_{2}$&------   &0.000 &----   &23341.495 &45.563   &1.0125 &23235.615 &134.784  &0.9984   &180\\
$3d^{2}$$^{3}P_{0}$&------   &0.000 &----   &33314.030 &110.231  &1.0021 &-----     &0.000    &----     &110\\
$3d^{2}$$^{3}P_{1}$&33379.722&36.971&1.0083 &33327.560 &27.594   &1.0034 &33221.680 &45.540   &0.9936   &110\\
$3d^{2}$$^{3}P_{2}$&------   &0.000 &----   &33353.459 &27.662
&1.0059 &33247.579 &82.178   &0.9961   &110\\\hline
$4s4d^{3}D_{1}$    &22590.296&48.981&1.0134 &22538.134 &36.471   &1.0061 &22432.254 &2.396    &0.9916   &87\\
$4s4d^{3}D_{2}$    &---------&0.000 &------ &22541.804 &65.687   &1.0067 &22435.924 &21.580   &0.9922   &87\\
$4s4d^{3}D_{3}$    &---------&0.000 &------ &--------- &0.000    &------ &22441.506 &86.400   &0.9931   &87\\
$4s5d^{3}D_{1}$    &27585.101&20.786&1.0112 &27532.939 &15.498   &1.0053 &27427.059 &1.021    &0.9934   &37\\
$4s5d^{3}D_{2}$    &---------&0.000 &------ &27534.653 &27.905   &1.0056 &27428.773 &9.192    &0.9937   &37\\
$4s5d^{3}D_{3}$    &---------&0.000 &------ &--------- &0.000    &------ &27431.444 &36.782   &0.9941   &37\\
$4s6d^{3}D_{1}$    &29891.172&13.472&1.0104 &29839.010 &10.049   &1.0049 &29733.130 &0.663    &0.9940   &24\\
$4s6d^{3}D_{2}$    &---------&0.000 &------ &29840.356 &18.094   &1.0052 &29734.476 &5.965    &0.9942   &24\\
$4s6d^{3}D_{3}$    &---------&0.000 &------ &--------- &0.000    &------ &29736.431 &23.868   &0.9945   &24\\
$4s7d^{3}D_{1}$    &31144.072&8.417 &1.0100 &31091.910 &6.279    &1.0047 &30986.030 &0.414    &0.9942   &15\\
$4s7d^{3}D_{2}$    &---------&0.000 &------ &31093.586 &11.306   &1.0050 &30987.706 &3.729    &0.9944   &15\\
$4s7d^{3}D_{3}$    &---------&0.000 &------ &--------- &0.000    &------ &30990.116 &14.922   &0.9948   &15\\
$4s8d^{3}D_{1}$    &31878.324&4.318 &1.0094 &31826.162 &3.221    &1.0041 &31720.282 &0.213    &0.9940   &7.7\\
$4s8d^{3}D_{2}$    &---------&0.000 &------ &31829.944 &5.803    &1.0048 &31724.064 &1.915    &0.9946   &7.7\\
$4s8d^{3}D_{3}$    &---------&0.000 &------ &--------- &0.000    &------ &31729.298 &7.663    &0.9952   &7.7\\
$4s3d^{3}D_{1}$    &5177.459 &0.502 &1.0503 &5125.297  &0.365    &1.0188 &5019.417  &0.023    &0.9570   &0.86$^{a}$\\
$4s3d^{3}D_{2}$    &---------&0.000 &------ &5139.197  &0.663    &1.0272 &5033.317  &0.207    &0.9650   &0.86$^{a}$\\
$4s3d^{3}D_{3}$    &---------&0.000 &------ &--------- &0.000
&------ &5055.057  &0.841    &0.9775   &0.86$^{a}$\\\hline
\multicolumn{11}{|l|}{\rule[1.5mm]{0mm}{3mm}\rule[-1mm]{0mm}{3mm}$^{a}$~\cite{Carsten}.}\\\hline
\end{tabular}
\end{widetext}
\end{scriptsize}
\end{tablehere}

between $A_{Tki}$ and $A_{Jki}$. We know $A_{Tki}$ can be
expressed as:
\begin{eqnarray}
A_{Tki} = \frac{e^2}{4\pi\epsilon_0}\frac{{4\omega
_{Tki}^3}}{{3\hbar {c^3}}} \frac{1}{{2{L_k} + 1}} |\left\langle
{{\beta _k}{L_k}}\left\| D \right\| {{\beta _i}{L_i}} \right\rangle|
^2
\end{eqnarray}
Here $\hbar\omega_{Tki}$ is the energy difference between two fine
structure manifold states $|k\rangle$ and $|i\rangle$. Using the
formula:
\begin{eqnarray}
\begin{split}
\left\langle {{\beta_k}{J_k}} \left\| D \right\|
{{\beta_i}{J_i}} \right\rangle
&= {( - 1)^{{L_k} + {S_k} + {J_i} + 1}}{\delta _{{S_k}{S_i}}}\sqrt {(2{J_k} + 1)(2{J_i} + 1)}\\
&\times \left\{ {\begin{array}{*{20}{c}}
   {{J_k}} & 1 & {{J_i}}  \\
   {{L_i}} & {{S_i}} & {{L_k}}  \\
\end{array}} \right\}\left\langle {{\beta _k}{L_k}} \left\| D \right\| {{\beta _i}{L_i}} \right\rangle\\
\end{split}
\end{eqnarray}
and combining Eq.(2) and (3), we can get
\begin{eqnarray}
A_{Jki}  = A_{Tki}\times \zeta (\omega _{ki} ){\widetilde {R_{ki}} }
\end{eqnarray}
Here
\begin{eqnarray}
\zeta (\omega _{ki} )={\omega _{Jki}^3}/\omega _{Tki}^3
\end{eqnarray}
is the energy dependent correction~\cite{Martin}, reflecting the alteration on the transition rate due to the effects
such as the orbit-spin interaction and the spin-spin interaction
which causes the fine structure splitting.
And
\begin{eqnarray}
\begin{split}
\widetilde {R_{ki}}&=(2{L_k} + 1)(2J_i + 1) \times {\left\{
{\begin{array}{*{20}{c}}
   J_k & 1 & J_i  \\
   L_i & S_i & L_k  \\
\end{array}} \right\}^2}
\end{split}
\end{eqnarray}
gives the fraction of the coupling strength between an excited state
$|k\rangle$ and a lower state $|i\rangle$. Since the total
transition rate $A_{Tki}$ is usually available in the literature,
this geometric ratio tells us how to scale the interaction for a
particular fine structure state of interest.

To calculate the wavelength dependent polarizability, we combine
Eq.(1) with Eq.(5), and use the known transition frequencies and
spontaneous emission rates in the literature. This light
polarizability is very sensitive to the Einstein coefficient.
However, theoretical and experimental values of magic wavelength for
the optical clock transition obtained in the past can be used to
confirm our calculation.

In this paper, we use this method to calculate the light shift for
the terahertz clock transition from $^{3}P_{0}$ to $^{3}P_{1},m=0$
levels, and from $^{3}P_{1},m=0$ to $^{3}P_{2},m=0$ levels for boson
isotopes with the nuclear spin $I=0$. After calculating magic
wavelengths for Sr and Ca  optical clock transitions and comparing
them to the experimental values, we calculate the polarizability of
terahertz  transition with data collection mainly from
Ref.~\cite{NIST, EMoore,SrA,Srb,Src,Srd,Sre,Srf,Srg}.

\section{Calculation of Magic wavelength }

\subsection{Strontium}

Using the method above, for Strontium, we first calculate the magic
wavelengths of two optical lattice clock transitions with the datas
listed in Table I and compare the results with experimental values.
Then, we calculate the crossing points for terahertz clock
transitions where the difference of polarizability is zero. Table I
shows Transition Wavenumbers, Einstein Coefficients and Correction
Factors for the $5s^{2}$~$^{1}S_{0}$, $5s5p~^{3}P_{0}$,
$5s5p~^{3}P_{1}$ and $5s5p~^{3}P_{2}$ states for Sr element. For
Einstein Coefficient $A_{Tki}$, first we choose the available
updated experimental values in Ref.~\cite{Srb,Srd,Srg}, then we use
updated theoretical data in Ref.~\cite{Src,Srf}, and for the rest we
mainly use theoretical values in Ref.~\cite{SrA}.

According to our calculation, the crossing point for the $^{1}S_{0}$
to $^{3}P_{0}$ transition occurs at 813.1 nm, while the crossing
point for the $^{1}S_{0}$ to $^{3}P_{1}(m_{J}={\pm}1)$ transition
with linear polarized light takes place at 915.4 nm. Both of those
results are in agreement with the experimental values of 813.428(1)
nm~\cite{Martin58, Martin44, Martin41, Martin42} and 914(1)
nm~\cite{Src}. This confirms our calculation procedure.

\begin{figure}[ht]
\includegraphics [width=9cm]{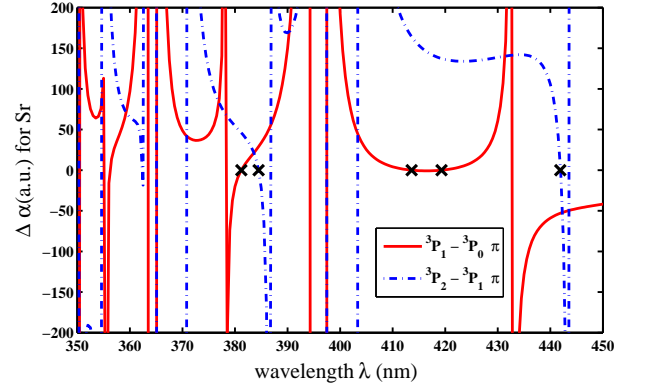}
\caption{(Color online) The wavelength dependence of the difference of atomic
polarizability for Sr element around 400 nm.}
\end{figure}

\begin{figure}[ht]
\includegraphics[width=9cm]{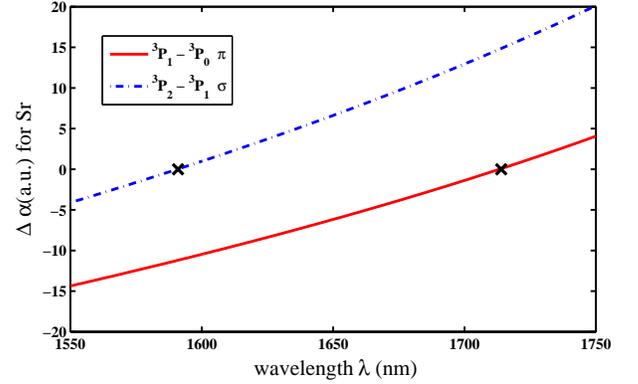}
\caption{(Color online) The wavelength dependence of the difference of atomic
polarizability for Sr element around 1650 nm.}
\end{figure}
Fig. 2 and Fig. 3 display the wavelength dependence of the atomic
polarizability difference $\Delta\alpha$ for Sr with trapping laser
wavelength around 400 nm and 1650 nm, respectively. The result is
scaled by a factor of 1/($4\pi \varepsilon _0 a_0^3$) and the
polarizability is given in atomic unit. In Fig.2, for linear
polarized light, $\Delta\alpha$ between $^3P_1$ and $^3P_0$ and
$\Delta\alpha$ between $^3P_2$ and $^3P_1$ are given in solid and
dash dotted lines, respectively. In Fig.3, $\Delta\alpha$ between
$^3P_1$ and $^3P_0$ for linear polarized light and $\Delta\alpha$
between $^3P_2$ and $^3P_1$ for circular polarized light are
presented. The cross markers are the crossing points where
$\Delta\alpha$ is zero. From Fig.2 and 3, we can know that the magic
wavelength for $^{3}P_{0}$ to $^{3}P_{1},m=0$ with linear polarized
light are 381.2 nm, 413.6 nm, 419.3 nm, 1714 nm and 3336 nm, while
for $^{3}P_{1},m=0$ to $^{3}P_{2},m=0$ are 384.5 nm,441.9 nm and
511.0 nm. On the other hand, for $m=0$ and circular polarization of
light, the magic wavelengths for $^{3}P_{0}$ to $^{3}P_{1},m=0$
transition are 511.8 nm and 662.8 nm, while for $^{3}P_{1},m=0 $ to
$^{3}P_{2},m=0$, the magic wavelengths are 717.7 nm and 1591 nm.

\subsection{Calcium}

We calculate the polarizabilities using the data in Table II with
the same method. Table II shows Transition Wavenumbers, Einstein
Coefficients and Correction Factors for the $4s^{2}~$$^{1}S_{0}$,
$4s4p~^{3}P_{0}$, $4s4p~^{3}P_{1}$ and $4s4p~^{3}P_{2}$ states for
Ca. For Einstein Coefficient, we use the updated theoretical values
according to Ref.~\cite{Carsten}, and others are from the data
listed in Ref.~\cite{NIST}. In order to check the accuracy of our
calculation and the data used, we get the magic wavelength 799.2 nm
for the $^{1}S_{0}, m=0$ to $^{3}P_{1}, m=0$ optical transition with
circularly polarized trapping light, which agrees well with the
experimental value 800.8( 22 ) nm~\cite{Carsten}.

The wavelength dependence of the atomic polarizability difference
$\Delta\alpha$ around 350 nm and 1350 nm are shown with atomic unit
in Fig.4 and 5, respectively. The crossing points where the
$\Delta\alpha$ is zero are marked by cross. The magic wavelengths
for linear polarization occur at 1361 nm and 2066 nm for clock
transition $^{3}P_{0}-^{3}P_{1}$, while for the transition between

\begin{tablehere}
\begin{scriptsize}
\begin{widetext} \caption{Mg element: Transition wavenumbers (WN)($cm^{-1}$)
corresponding to $\omega_{Jki}$, Einstein Coefficients for fine
structure states $A_J(\times10^6s^{-1})$ and Total
$A_T(\times10^6s^{-1})$ for fine structure states manifold,
Correction Factors $\boldsymbol{\zeta}$ for Mg element. The
Wavenumber and $A_T$ data originate from~\cite{NIST}.}
\begin{tabular}{|c|c|c|c|c|}\hline
\hline{$$}&\multicolumn{3}{c|}{{$3s^{2}$$^{1}S_{0}$}}&{$$}\\\hline
$\textbf{\textrm{UpperState}}$&{WN}&{${A_{J}}$}&{$\boldsymbol{\zeta}$}&\raisebox{0ex}[0pt]{${A_{T}}$}\\\hline
$3s3p^{1}P_{1}$    &35051.264&491  &1.000                &491\\
$3s4p^{1}P_{1}$    &49346.729&61.2 &1.000                &61.2\\
$3s5p^{1}P_{1}$    &54706.536&16.0 &1.000                &16.0\\
$3s6p^{1}P_{1}$    &57214.990&6.62 &1.000                &6.62\\
$3s7p^{1}P_{1}$    &58580.230&3.28 &1.000                &3.28\\
$3s8p^{1}P_{1}$    &59403.180&1.88 &1.000                &1.88\\
\hline
\end{tabular}
\\
\begin{tabular}{|c|c|c|c|c|c|c|c|c|c|c|}\hline
\hline{$$}&\multicolumn{3}{c|}{{$3s3p^{3}P_{0}$}}
&\multicolumn{3}{c|}{{$3s3p^{3}P_{1}$}}
&\multicolumn{3}{c|}{{$3s3p^{3}P_{2}$}} &{$$}\\\hline
$\textbf{\textrm{UpperState}}$&{WN}
&{${A_{J}}$}&{$\boldsymbol{\zeta}$}&{WN}&{${A_{J}}$}&{$\boldsymbol{\zeta}$}&{WN}
&{${A_{J}}$}&{$\boldsymbol{\zeta}$}&\raisebox{0ex}[0pt]{${A_{T}}$}\\\hline
$3s4s^{3}S_{1}$    &19346.998&11.293    &1.0063 &19326.939  &33.774      &1.0032 &19286.225 &55.932   &0.9968   &101\\
$3s5s^{3}S_{1}$    &30022.121&3.380     &1.0041 &30002.062  &10.120      &1.0020 &29961.348 &16.800   &0.9980   &30.3\\
$3s6s^{3}S_{1}$    &34041.395&1.372     &1.0036 &34021.336  &4.107
&1.0018 &33980.622 &6.821    &0.9982   &12.3\\\hline
$3p^{2}$$^{3}P_{0}$&-------  &0.000     &----   &35942.306  &537.085   &0.9983 &-------   &0.000    &----     &538\\
$3p^{2}$$^{3}P_{1}$&35982.995&179.638 &1.0017 &35962.936  &134.500   &1.0000 &35922.222 &223.405  &0.9966   &538\\
$3p^{2}$$^{3}P_{2}$&-------  &0.000     &----   &36003.476  &134.957
&1.0034 &35962.762 &403.500  &1.0000   &538\\\hline
$3s3d^{3}D_{1}$    &26106.653&89.865    &1.0047 &26086.594  &67.244      &1.0024 &26045.880 &4.462    &0.9977   &161\\
$3s3d^{3}D_{2}$    &-------  &0.000     &----   &26086.563  &121.028   &1.0023 &26045.849 &40.157   &0.9977   &161\\
$3s3d^{3}D_{3}$    &-------  &0.000     &----   &-------    &0.000     &----   &26045.867 &160.630  &0.9977   &161\\\
$3s4d^{3}D_{1}$    &32341.930&28.106    &1.0038 &32321.871  &21.040      &1.0019 &32281.157 &1.397    &0.9981   &50.4\\
$3s4d^{3}D_{2}$    &-------  &0.000     &----   &32321.830  &37.872      &1.0019 &32281.116 &12.576   &0.9981   &50.4\\
$3s4d^{3}D_{3}$    &-------  &0.000     &----   &-------    &0.000     &----   &32281.078 &50.304   &0.9981   &50.4\\
$3s5d^{3}D_{1}$    &35117.866&13.101    &1.0035 &35097.807  &9.808     &1.0017 &35057.093 &0.652    &0.9983   &23.5\\
$3s5d^{3}D_{2}$    &-------  &0.000     &----   &35097.784  &17.655      &1.0017 &35057.070 &5.865    &0.9983   &23.5\\
$3s5d^{3}D_{3}$    &-------  &0.000     &----   &-------    &0.000     &----   &35057.040 &23.460   &0.9983   &23.5\\
$3s6d^{3}D_{1}$    &36592.473&6.967     &1.0033 &36572.414  &5.217     &1.0017 &36531.700 &0.347    &0.9983   &12.5\\
$3s6d^{3}D_{2}$    &-------  &0.000     &----   &36572.381  &9.391     &1.0017 &36531.660 &3.120    &0.9983   &12.5\\
$3s6d^{3}D_{3}$    &-------  &0.000     &----   &-------    &0.000
&----   &36531.657 &12.479   &0.9983   &12.5\\\hline
\end{tabular}
\end{widetext}
\end{scriptsize}
\end{tablehere}

level $^{3}P_{1}, m=0$ and $^{3}P_{2},m=0 $ at 312.2 nm, 316.2 nm,
325.4 nm, 344.0 nm and 393.4 nm.

The laser polarization have no effect on the polarizability for the
ground state ( $J=0$ ) because the ac Stark shift is identical with
any polarizations. It is also true for $^{3}P_{0}$ state. However,
the influence of circular polarized laser light is worth study for
other states. For $m=0$, we can obtain the magic wavelengths for the
$^{3}P_{0}$ to $^{3}P_{1}$ clock transition at 301.0 nm and 310.0
nm, while for $^{3}P_{1}$ to $^{3}P_{2}$ one finds 1318 nm and 2254
nm.

\begin{figure} \centering
\includegraphics[width=9cm]{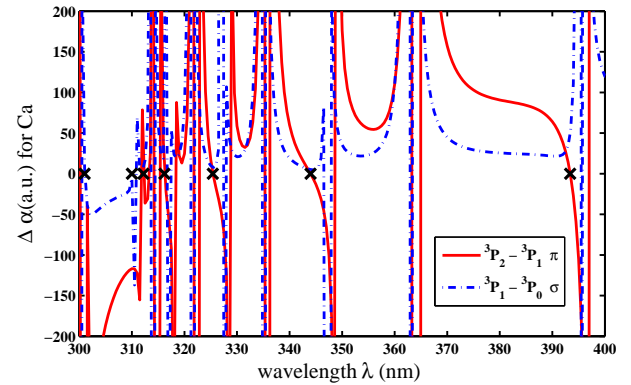}
\caption{(Color online) Wavelength dependence of the difference between excited state and ground state atomic
polarizability around 350 nm for Ca element.}
\end{figure}

\begin{figure} \centering
\includegraphics[width=9cm]{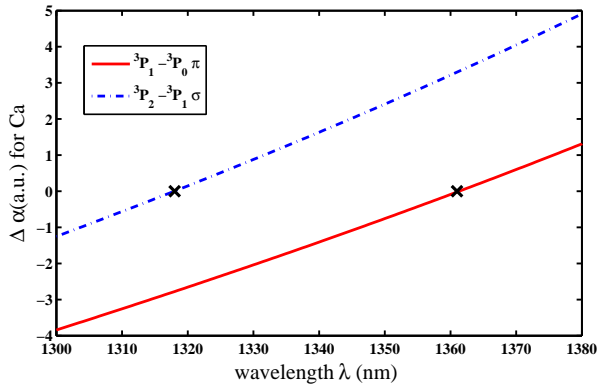}
\caption{(Color online) Wavelength dependence of the difference between excited state and ground state atomic
polarizability around 1350 nm for Ca element.}
\end{figure}

\subsection{Magnesium}

With the completion of the NIST database, the atomic polarizability
of the Mg triplet states in the presence of linear and circular
polarized light can also be calculated. Table III presents
Transition Wavenumbers, Einstein Coefficients and Correction Factors
for the $3s^{2}~$$^{1}S_{0}$, $3s3p~^{3}P_{0}$, $3s3p~^{3}P_{1}$ and
$3s3p~^{3}P_{2}$ states for Mg element. Using the data presented in
Table III, the magic wavelengths of $^{3}P_{0}$ to $^{3}P_{1},m=0$
transition with linear polarization are 335.6 nm and 399.5 nm. The
magic wavelengths are 308.6 nm, 336.5 nm, 406.1 nm for the
transition between $^{3}P_{1},m=0$ and $^{3}P_{2},m=0$.

For circular polarization of light, the magic wavelengths for the
transition $^{3}P_{0}$ to $^{3}P_{1},m=0$ are 307.7 nm, 336.4 nm,
407.8 nm. However, we can not find any magic wavelength for circular
laser between level $^{3}P_{1},m=0 $ and $^{3}P_{2},m=0$.

For Mg atoms, several optical transitions between the energy levels
of terahertz clock transition states and other levels exist, such as
456.5 nm ( ${3s^{2}}~{^1S} - 3s3p~^{3}P$ ), 383.6 nm ( $3s3p~^{3}P -
3s3d~^{3}D$ ), 309.6 nm ( $3s3p~^{3}P - 3s4d~^{3}D$ ), 333.2 nm (
$3s3p~^{3}P - 3s5s~ ^{3}S$ ) and 517.4 nm ( $3s3p~^{3}P -
3s4s~^{3}S$ ). Hence, not all the magic wavelengths are good enough
for clock transition, because the slope of the light shift
difference with the wavelength is too large (shown in the final
table IV). To some extent, a possible magic wavelength near 400 nm
is shown in Fig. 6 with the atomic unit. In Fig.6, $\Delta\alpha$
for $^3P_1-^3P_0$ transition and $^3P_2-^3P_1$ transition with
different polarization are given. The cross markers reflect the
crossing points where the atomic polarizability difference is zero.

\begin{figure} \centering
\includegraphics[width=9cm]{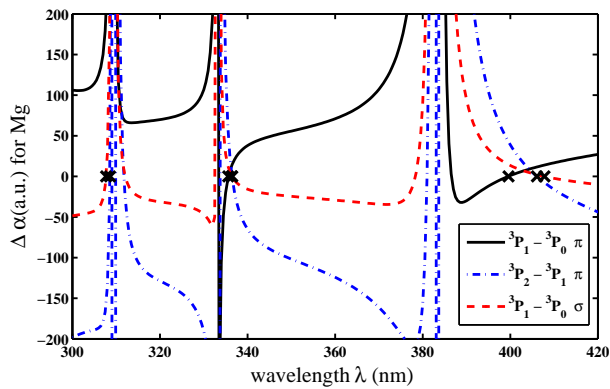}
\caption{(Color online) The wavelength dependence of the difference of atomic
Polarizability around 400 nm for Mg element.}
\end{figure}

\section{Discussions and Conclusions}

\begin{table}[ht]
\begin{flushleft}
\caption{Magic wavelengths for terahertz region. L1 is the linear
laser for the $^{3}P_{0}$ to $^{3}P_{1}$ clock transition while C1
is for the circular laser, L2 is the linear laser for the
$^{3}P_{1}$ to $^{3}P_{2}$ clock transition while C2 is for the
circular laser. $\kappa$ is the slope of the shift difference of two
level of clock transition levels at the corresponding trapping laser
wavelength with unit $Hz/nm$, the sign denotes the direction of the
change with the shift for the high level minus the lower level. The
data are given in the reasonable experiment condition with input
power 150 $mW$ focused to a waist of 65 $\mu m$ as the light
intensity $1.1301*10^3 W/cm^2$.}
\end{flushleft}
\centering
\begin{tabular}{|c|c|c|c|c|c|c|}\hline\hline
 &\multicolumn{2}{c|}{$Mg$}&\multicolumn{2}{c|}
 {\rule[1.5em]{2.5mm}{0mm}\rule[-0.8em]{2.5mm}{0mm}$Ca$\rule[1.5em]{5mm}{0mm}}&\multicolumn{2}{c|} {$Sr$}\\\cline{2-7}
 \raisebox{1.5ex}[0pt]{$\boldsymbol{^{3}P_{0,1,2}}$}&\rule[1.5em]{4mm}{0mm}$\lambda$\rule[1.5em]{4mm}{0mm}&$\kappa$&
 \rule[1.5em]{4mm}{0mm}$\lambda$\rule[1.5em]{4mm}{0mm}&$\kappa$&\rule[1.5em]{4mm}{0mm}$\lambda$\rule[1.5em]{4mm}{0mm}&$\kappa$\\
 &$(nm)$\rule[-1em]{0mm}{0mm}&($Hz/nm$)&$(nm)$&($Hz/nm$)&$(nm)$&($Hz/nm$)\\
 \hline
 \rule[1mm]{0mm}{3mm}\rule[-1.5mm]{0mm}{4mm}$L1$       &335.6&-1125&1361&-3.201&381.2&-615.7\\
 \cline{2-7}\rule[1mm]{0mm}{3mm}\rule[-1.5mm]{0mm}{4mm}&399.5&-103.5&2066&54.94&413.6&32.22\\ \cline{2-7}
  \rule[1mm]{0mm}{3mm}\rule[-1.5mm]{0mm}{4mm}          &&&&&419.3&-31.4377\\ \cline{2-7}
  \rule[1mm]{0mm}{3mm}\rule[-1.5mm]{0mm}{4mm}          &&&&&1714&-5.590\\ \cline{2-7}
  \rule[1mm]{0mm}{3mm}\rule[-1.5mm]{0mm}{4mm}          &&&&&3336&9.122\\
  \hline
 \rule[1mm]{0mm}{3mm}\rule[-1.5mm]{0mm}{4mm}$L 2$      &308.6&-26574&312.2&8319&384.5&1792\\ \cline{2-7}
\rule[1mm]{0mm}{3mm}\rule[-1.5mm]{0mm}{4mm}
&336.5&1905&316.2&5879&441.9&5759\\\cline{2-7}
\rule[1mm]{0mm}{3mm}\rule[-1.5mm]{0mm}{4mm}
&406.1&220.7&325.4&1641&511&1697.63\\ \cline{2-7}
 \rule[1mm]{0mm}{3mm}\rule[-1.5mm]{0mm}{4mm}           &&&344.0&542.9&&\\\cline{2-7}
 \rule[1mm]{0mm}{3mm}\rule[-1.5mm]{0mm}{4mm}           &&&393.4&1787&&\\\cline{2-7}
\hline
 \rule[1mm]{0mm}{3mm}\rule[-1.5mm]{0mm}{4mm}$C1$       &307.7&-3174&301.0&8610&511.8&252.8\\ \cline{2-7}
  \rule[1mm]{0mm}{3mm}\rule[-1.5mm]{0mm}{4mm}          &336.4&469.9&310.0&-4072&662.8&-1068\\ \cline{2-7}
  \rule[1mm]{0mm}{3mm}\rule[-1.5mm]{0mm}{4mm}          &407.8&54.76&&&&\\
  \hline
 \rule[1mm]{0mm}{3mm}\rule[-1.5mm]{0mm}{4mm}$C 2$      &&&1318&-3.365&717.7&2692\\ \cline{2-7}
 \rule[1mm]{0mm}{3mm}\rule[-1.5mm]{0mm}{4mm}           &&&2254&13.39&1591&-5.551\\ \hline
\end{tabular}
\end{table}

In summary, we have calculated magic wavelengths for terahertz clock
transitions for alkaline-earth atoms. The calculation results are
presented in Table IV along with the slopes of the difference of
polarizabilities at corresponding magic wavelengths. Depending on
the calculation and the current laser development, we recommend
1714nm and 1591nm for Sr terahertz clock, 1361nm and 1318nm for Ca
terahertz clock, 399.5nm and 407.8nm for Mg terahertz clock, because
the difference of polarizabilities have small slopes at these magic
wavelengths, where we ignore the effect of highly excited states and
continuum states which can only make little contribution to the
wavelength dependent polarizabilities at terahertz region.

In this paper, we are only focusing on the study of possible magic
wavelengths of trapping laser for these terahertz clock transitions
of Sr, Ca and Mg atoms. These terahertz clock transitions were first
proposed as early as 1972~\cite{Strumia}, and recently have been
proposed to be applied in active optical clock~\cite{Chen}. These
clock transitions of alkaline-earth atoms correspond to a 0.6 THz to
11.8 THz frequency region. After the successful developments of
microwave fountain frequency standards, optical clocks with trapped
ions and optical lattice trapped neutral atoms, it is interesting to
study clock transitions at terahertz wavelengths. The advantages and
disadvantages of terahertz magic atomic clock will be discussed
elsewhere. The wavelength range studied in this paper (from 500
$\mu$m to 25 $\mu$m) corresponding to THz frequency standards fills
the gap between microwaves and optical waves.

We thank V. Thibault for critical reading our manuscript, and J. M. Li for his helpful discussions.
We appreciate the anonymous referee for the useful suggestions. This work is partially supported by the state Key Development Program for
Basic Research of China (No.2005CB724503, 2006CB921401,
2006CB921402) and NSFC (No.10874008, 10934010, 60490280 and
10874009). This work is also supported by PKIP of CAS (KJCX2.YW.W10) and Open Research Found of
State Key Laboratory of Precision Spectroscopy (East China
Normal University).\\

\end{document}